\documentstyle[11pt,newpasp,twoside,epsf]{article}
\def\etal{{et\thinspace al.}\ }
\def\Teff{$T_{\sf eff}$}
\def\logg{$\log g$\,}
\def\mdot{$\dot{M}$}

\begin{document}
\title{Stellar Atmosphere and Accretion Disk Models for the Hot Component in
Symbiotic Stars}
\author{Klaus Werner$^1$, Jochen L. Deetjen$^1$, Stefan Dreizler$^1$, Thorsten Nagel$^1$,
        Thomas Rauch$^{2,1}$}
\affil{$^1$Institut f\"ur Astronomie und Astrophysik, Universit\"at T\"ubingen, 
Sand 1, 72076 T\"ubingen, Germany}
\affil{$^2$Dr. Remeis-Sternwarte, Universit\"at Erlangen-N\"urnberg, Sternwartstr.
7, 96049 Bamberg, Germany}

\begin{abstract}
We describe our NLTE codes which allow the computation of synthetic spectra of
hot stars and accretion disks. They can be combined to compute ionizing fluxes
from the hot component in symbiotic stars.
\end{abstract}

\section{Introduction}

Symbiotic stars are interacting binaries with a relatively large separation when
compared to cataclysmic variables. Their orbital periods are of the order of
months up to many decades. Symbiotic binaries consist of a cool giant star and
a hot ionizing radiation source (see e.g.\ Kenyon 2001). The giant looses mass
by a wind and not by Roche-lobe overflow (M\"urset \& Schmid 1999) at a rate
of typically $10^{-6}$\,M$_{\odot} {\rm yr}^{-1}$. Only a few percent of this
amount is accreted by the companion. The hot component (in quiescent symbiotics)
is in most cases a hot white dwarf. Two known systems are hard X-ray sources and
contain accreting neutron stars (Chakrabarty \& Roche 1997, Masetti \etal
2002). It is not clear if an accretion disk is present in all systems. The
possible formation of stable disks is supported by hydrodynamic simulations (Mastrodemos
\& Morris 1998, Dumm \etal 2000).

The observed high luminosity of the hot component is generated by mass accretion from
the giant wind. An accretion rate of \mdot\,$\approx10^{-7}\,$M$_{\odot}
{\rm yr}^{-1}$ onto a white dwarf yields a luminosity of 100 L$_{\odot}$. It is
believed that hydrogen shell burning of the accreted material can be sustained
above the white dwarf core and in this case \mdot\,$\approx10^{-8}\,$M$_{\odot}
{\rm yr}^{-1}$ is sufficient to generate that amount of luminosity.

The white dwarf is dominating the hot component in those symbiotics where steep
UV continua are observed. The derived white dwarf parameters are $R \approx
0.1\,$R$_{\odot}$, $M$\,=\,0.5--1 M$_{\odot}$, and the effective temperatures
range between \Teff\,=\,30\,000 up to 200\,000\,K (see e.g.\ M\"urset \etal
1991). The accretion disk is probably dominant in about 10\% of the systems
with flat UV continua (Kenyon 2001). The disk temperature reaches 100\,000\,K
at the inner boundary layer and decreases outwards.

It is obvious that the analysis of emission line spectra from symbiotic stars
requires a realistic modeling of the white dwarf and the disk. Occasionally, the
collision of winds from the white dwarf and the disk as well as jets can
contribute significantly to the ionizing radiation. In the following we present
our approach to calculate synthetic spectra of the hot component. In Sect.\,2.\
we describe the construction of white dwarf and neutron star NLTE model
atmospheres, assuming planar geometry and hydrostatic equilibrium. Then we
present our method to compute NLTE disk models, which assume a radial Keplerian
$\alpha$-disk structure and a detailed vertical structure (Sect.\,3.). Up to
now we are neglecting stellar winds and disk winds. We also disregard jets
which are observed in some symbiotics and which could result from outbursts,
disk instabilities, or very high accretion rates.

\section{Stellar Atmosphere Models}

Our computer code calculates plane-parallel, hydrostatic NLTE models in
radiative equilibrium. It uses an Accelerated Lambda Iteration (ALI) technique
and it is described in detail by Werner \& Dreizler (1999) and Werner \etal
(2002). The code solves self-consistently the equations of radiation transfer,
hydrostatic and radiative equilibrium, particle conservation, and the NLTE rate
equations for the atomic levels populations. Full NLTE line blanketing is
possible. Iron group elements are included by a statistical approach
(superlevels and -lines) and an opacity sampling method. Basic input parameters
are \Teff, \logg, and the chemical composition. The code computes the
atmospheric structure and an emergent spectrum. It is routinely used
for the spectral analysis in the X-ray to IR spectral regions of many objects:
Central stars of planetary nebulae, white dwarfs, and subdwarfs. For white
dwarfs element diffusion strongly affects the (E)UV flux, however, this is
unimportant for accreting white dwarfs because of the short diffusion time scales.
This process could only play a role when accretion were shut off over a period
of many years. We have successfully computed and applied such models (Dreizler
\& Wolff 1999, Schuh \etal 2002). Some symbiotic systems harbor a magnetic
white dwarf. In these cases as well as in the systems with neutron star
components the magnetic field affects the emergent stellar spectrum. A polarized
radiation transfer calculation is possible with an extension of our code
(Deetjen \etal 2002). We already applied our code to the analysis of UV
and X-ray data from symbiotic stars (Jordan \etal 1994, 1996).

An extensive grid of stellar fluxes, also relevant for white dwarfs in
symbiotic systems, is available from {\tt
http://astro.uni-tuebingen.de/\~\,rauch/} (Rauch 2002). The data have been
calculated for photoionization modeling, in particular with the {\sc CLOUDY}
code (Ferland \etal 1998).

\section{NLTE Accretion Disk Models}

A new computer code, that has been developed from the stellar atmosphere code
described above, calculates disk spectra under the following assumptions (for
more details see Nagel \etal (2002) and in this volume). The radial disk structure is
computed assuming a stationary, Keplerian, geometrically thin $\alpha$-disk
(Shakura \& Sunyaev 1973). This model is fixed by four global input parameters:
Stellar mass $M_\star$ and radius $R_\star$ of the accretor, mass accretion
rate \mdot, and the viscosity parameter $\alpha$. For numerical treatment the
disk is divided into a number of concentric rings. For each ring with radius $R$
our code calculates the detailed vertical structure, assuming a plane-parallel
radiating slab (in analogy to a stellar atmosphere).

In contrast to a (planar) stellar atmosphere, which is characterized by \Teff\
and \logg, a particular disk ring with radius $R$ is characterized by the
following two parameters, which follow from the global disk parameters
introduced above. The first parameter measures the dissipated and then radiated
energy. It can be expressed in terms of an effective temperature \Teff:
$$T_{\rm eff}^4(R)=3GM_\star\dot{M}/8\sigma\pi R^3 \cdot (1-R_\star/R)^{1/2}.$$
The second parameter is the surface mass density $M$ of the disk ring: $$
M(R)=[1-(R_\star/R)^{1/2}]\, \dot{M}/3\pi \bar{w}.$$ $\sigma$ and $G$ are the
Stefan-Boltzmann and gravitational constants, respectively. $\bar{w}$ is the
depth mean of viscosity $w(z)$, where $z$ is the height above the disk
mid-plane. The viscosity is given by the standard $\alpha$-parametrization as a
function of the total (i.e.\ gas plus radiation) pressure, but numerous other
modified versions are used in the literature. We use a formulation involving
the Reynolds number, as proposed by Kriz \& Hubeny (1986) and Hubeny \& Hubeny
(1998).

The radiation transfer equations plus vertical structure equations are then
solved like in the stellar atmosphere case, but accounting for two basic
differences. First, the gravitational acceleration (entering the hydrostatic
equation) is not constant with depth, but increases with $z$. This is simply
the vertical component of the gravitational acceleration exerted by the central
object (self-gravitation of the disk is negligible):
$$g=z\, GM_\star/R^3.$$
Second, the energy equation for radiative equilibrium balances the dissipated
mechanical energy and the net radiative losses:
$$ 9/4\,\, \rho w\, GM/R^3 = 4\pi
\int_{0}^{\infty}(\eta_\nu-\kappa_\nu J_\nu)d\nu ,$$
where $\rho$, $\eta$, $\kappa$, $J$ denote mass density, opacity, emissivity and mean 
intensity, respectively. In the case of a stellar atmosphere the
left-hand side of this equation vanishes and we get the usual radiative 
equilibrium equation.

The total observed disk spectrum, which depends on the inclination angle, is
finally obtained by intensity integration over all rings accounting for
rotational Doppler effects.

\section{Summary and Outlook}

We have modeling tools at hand which can be used to calculate ionizing spectra
of the hot component in symbiotic stars. We can compute synthetic spectra
emerging from the white dwarf (or neutron star) and from the accretion disk.
Our immediate aim for the near future is the inclusion of irradiation effects
of the stellar spectrum onto the disk. We also want to include effects of a
disk wind onto the emergent spectrum. Wind models for the compact star 
are already available (see e.g.\ Jordan \etal 1996) and must
be utilized at least for analyses of those symbiotic systems where hot stellar
winds are observed (Schmid 2000).

\begin{acknowledgements}
K. Werner thanks H.M. Schmid for helpful discussions. This work is supported by the
Deutsche Forschungsgemeinschaft (DFG) and the Deutsches Zentrum f\"ur Luft- und
Raumfahrt (DLR).
\end{acknowledgements}

\end{document}